\documentstyle[aps,graphicx]{revtex}
\def\vev#1{\left\langle #1\right\rangle}

\def\Im{\mathop{\mbox{Im}}}
\def\Re{\mathop{\mbox{Re}}}

\def\etal{{\it et al.}}

\def\qq{$\vev{\bar q q}$}
\def\GG{$\vev{\alpha_s GG/ \pi}$}
\def\eps{$\varepsilon$}
\def\eprime{$\varepsilon'$}
\def\ratio{$\varepsilon'/\varepsilon$}
\def\CP{$CP$}
\newcommand{\be}{\begin{equation}}
\newcommand{\ee}{\end{equation}}
\newcommand{\bea}{\begin{eqnarray}}
\newcommand{\eea}{\end{eqnarray}}

\begin{document}
\title{Estimating \ratio\ in the  standard model \\
independently of $\Im \lambda_t$}

\author{Marco Fabbrichesi}

\address{INFN, Sezione di Trieste and\\
Scuola Internazionale Superiore di Studi Avanzati\\
via Beirut 4, I-34013 Trieste, Italy.}

\date{\today}

\maketitle

\begin{abstract}
The $CP$-violating parameter \ratio\ is estimated in a novel way
by including the  explicit computation of \eps\ in the ratio as opposed to the
usual procedure of taking its value from the experiments. 
 This approach has the advantage of
being independent from the determination of the 
CKM parameters $\Im \lambda_t$ and of showing
more directly the dependence on the long-distance parameters.
The matrix elements are taken from the chiral quark model approach,
the parameters of which have been recently determined by 
an updated fit of the $CP$-conserving amplitudes entering the
$\Delta I=1/2$ rule.
By sampling the ranges of the experimental inputs according to
a normal distribution  and those of  the theoretical uncertainties
according to a flat one, it is found that
$$\varepsilon '/\varepsilon = (2.5 \pm 0.9) \times 10^{-3}\, ,$$
in agreement with the current experimental data. A more
conservative estimate
 $$0.9 \times 10^{-3} < \varepsilon '/\varepsilon < 5.8 \times 10^{-3}$$
is found  by considering the smallest and the largest values obtained by
taking a flat distribution for all inputs.
\end{abstract}




In the standard model,  the $CP$-violating parameter
\ratio\ can be in principle different from zero
because the $3 \times 3$ Cabibbo-Kobayashi-Maskawa (CKM) 
matrix $V_{ij}$, which appears in the weak charged currents
of the quark mass eigenstate, is in general complex.
On the other
hand, in other models like the superweak theory~\cite{W},
the only source of \CP\ violation resides in
the  $K^0$-$\bar K^0$ mixing, and \eprime\ vanishes. 
Establishing the precise value of \eprime\
is therefore of great importance.

Experimentally \ratio\ is extracted, by collecting $K_L$ and
$K_S$ decays into pairs of $\pi^0$ and $\pi^\pm$, from the relation 
\be
\Re\: \varepsilon '/\varepsilon \simeq
\left[ \left|\eta_{+-}/\eta_{00}\right|^2 -1 \right] /6 \, ,
\ee
and the determination of $\eta_{00}$ and $\eta_{+-}$ which are,
respectively, the ratio of the amplitudes for  
$K_L \rightarrow \pi^0 \pi^0$ over 
$K_S \rightarrow \pi^0 \pi^0$ and  
$K_L \rightarrow \pi^+ \pi^-$ over  
$K_S \rightarrow  \pi^+ \pi^-$.

With the announcement last year of the preliminary
result from the KTeV collaboration (FNAL)~\cite{KTeV}
based on data collected in 1996-97,
and from the NA48 collaboration (CERN)~\cite{NA48}
based on data collected in 1997-98,
the long-standing issue of whether \eprime\ vanishes or not seems to
be settled.
By computing the average (see Fig.~3 below) among the two 1992
experiments (NA31~\cite{NA31} and E731~\cite{E731}) 
and the preliminary data of KTeV and NA48 one obtains
\be
\Re\: \varepsilon/\varepsilon' = (1.9 \pm 0.46) \times 10^{-3}\, . 
\label{average}
\ee
The error in (\ref{average})
 has been inflated according to the Particle Data Group procedure for
combining results with substantially different central values.
The value and range in (\ref{average}) can be considered 
the current experimental result. Such a result
will be further improved by the complete run and full data analysis
from KTeV and NA48 and 
 the first data from KLOE (Frascati) thus achieving 
an uncertainty of only few parts in $10^{4}$.

From the theoretical point of view,
\eprime, which parameterizes direct \CP\ violation in the decays
 (for a review see, e.g., 
\cite{WW,burasrev,review}), is computed as
\be
\varepsilon ' = 
\frac{G_F \omega}{2\Re{A_0}} \:
\mbox{Im}\, \lambda_t \: \:
 \left[ \Pi_0 - \frac{1}{\omega} \: \Pi_2 \right] \; ,
\label{eprime}
 \ee
where $\Im \lambda_t \equiv \Im V_{td}V^*_{ts}$, 
\bea
 \Pi_0 & = &  
\frac{1}{\cos\delta_0} \sum_i y_i \, \Re\,  \langle  Q_i  \rangle _0 
\ (1 - \Omega_{\eta +\eta'}) \; ,
\label{Pi_0} \\
 \Pi_2 & = & 
\frac{1}{\cos\delta_2} \sum_i y_i \, \Re\, \langle Q_i \rangle_2 \; ,
\label{Pi_2}
\eea
and $\langle  Q_i  \rangle _I =  \langle 2 \pi , I|  Q_i| K
\rangle$. The phases $\delta_{0,2}$ come from final state
interactions. 
The explicit forms of $\Delta S=1$ four-quark operators
 $Q_i$ can be found, for instance, in \cite{review}.
The term $\Omega_{\eta+\eta'}$ is the isospin breaking (for $m_u\neq
m_d$)  contribution of the mixing of $\pi$ with $\eta$ and $\eta'$.
In (\ref{eprime}), the phase  
$\phi =  \pi/2 + \delta_0 - \delta_2 - \theta_\epsilon = (0\pm 4)^0$
has been taken as vanishing~\cite{Maiani}, and 
$CPT$ is assumed to hold. The Wilson coefficients $y_i$ are known to
the next-to-leading (NLO) order in $\alpha_s$ and $\alpha_w$~\cite{NLO}.
Notice the explicit final-state-interaction phases $\delta_I$
which comes from 
writing in (\ref{Pi_0}) and (\ref{Pi_2}) the absolute values 
of the amplitudes in term of their dispersive parts.

The parameter \eps, which parameterizes indirect \CP\ violation, is given by
\be
\varepsilon = \Im \lambda_t C_\varepsilon \hat B_K \left\{
\Re \lambda_c \left[ \eta_1 S_0(x_c) - \eta_3 S_1(x_c,x_t) \right]
- \Re \lambda_t \eta_2 S_0(x_t) \right\} \label{eps}
\ee
where $\lambda_c = V_{cd}V^*_{cs}$, $S_i(x)$ are the Inami-Lim 
functions~\cite{IL},
$x_i = m_i^2/m_W^2$ and 
$C_\varepsilon$ is a constant equal to 
$G_F^2 f_K^2 m_K m_W^2/(3 \sqrt{2} \pi^2 \Delta m_{LS})$.
$\hat B_K$ is the bag parameter to be determined in the long-distance
estimate of the hadronic matrix element for the $\Delta S=2$
transition. In (\ref{eps}), a term 
of higher order in $\lambda \equiv V_{us}$ is neglected. 
The parameters $\eta_i$, which
encode the renormalization group running, 
are known to the NLO order~\cite{s2}. 

In the usual approach---followed by all current estimates---the 
ratio \ratio\ is computed by dividing (\ref{eprime})
by the experimental value of \eps.
The allowed values for the CKM combination 
$\Im \lambda_t$ in front of  (\ref{eprime})
are then either taken from independent analysis of  the unitarity
triangle~\cite{dortmund,dubna,taipei}  or
consistently computed by means of
a determination of \eps\ in the same model (which includes the long-distance
matrix element of the $\Delta S=2$ 
amplitudes)~\cite{munich99,roma99,trieste97}.

This two-step procedure can be by-passed by computing 
in a given model \eprime\ and \eps\ and by taking directly 
their ratio. Because both
parameters are proportional to $\Im \lambda_t$, this quantity, which
is an important source of uncertainty, simplifies in the
ratio. This approach has also been advocated in~\cite{book,munich99}
without, however, any numerical analysis.
Of course, one trades the dependence on the Wolfenstein
parameter $\eta$ for that on $\rho$ in the factor 
$\Re \lambda_t = - \lambda (1-\lambda^2/2) V_{cb}^2 ( 1 -\bar \rho)$
that enters in (\ref{eps}). However, this dependence is weaker and
$\bar \rho \equiv (1 -\lambda^2/2) \rho$
can be determined reasonably well 
in the unitarity triangle independently of kaon physics.  
Another advantage of this procedure is that the dependence of
the final result on many of the input variables is made more
transparent
because the computation is not separated into the
determination of $\Im \lambda_t$ on the one hand and \ratio\ on the
other. However, as we shall see, the uncertainty in the final estimate
is about the same size as that obtained by the usual method, and it is
still dominated by the uncertainty in the hadronic matrix elements.

In this work I use this second approach and apply it to the estimate
of \ratio\ in the standard model.

The hadronic matrix elements
for all the relevant quark operators $Q_{1-10}$ 
and the parameter $\hat B_K$ are
computed in the chiral quark model ($\chi$QM)~\cite{QM,pich} 
at $O(p^4)$ in the chiral
expansion along the lines presented in~\cite{trieste97}.
I use this approach because it is the only theoretical
model in good agreement with the experimental data.

As we shall see, the final result is consistent with that of the
 $\chi$QM obtained by the usual method.
 
In the $\chi$QM approach, there are three model-dependent
parameters $M$, \qq\ $\equiv (\langle \bar q q \rangle)^{1/3}$
 and \GG\ which are fixed by means of a fit of
the $CP$-conserving amplitudes in
the $\Delta I = 1/2$ selection rule of $K\to\pi\pi$ decays.
This fit has been re-done recently in \cite{nuovo} by updating those
short-distance inputs that have become better known in the meantime.
The matching of the hadronic matrix elements to the
Wilson coefficients is done at the scale $\mu = 0.8$ GeV.
The  amplitudes $A(K^0 - 2 \pi, I=0)$
and  $A(K^0 - 2 \pi, I=2)$ computed in the model are compared with
their experimental values by allowing
at most a $\pm$ 20\% uncertainty. The values for  $M$, \qq\ and \GG\
found are those in Table~I.
The values for the two condensates vary according to the
 $\gamma_5$-scheme of dimensional regularization, 't Hooft-Veltman
 (HV) or Naive Dimensional Regularization (NDR), used in the
computation of the NLO Wilson coefficients and hadronic matrix
 elements. 

Having thus determined the model-dependent parameters, it is
possible to compute \ratio.

Notice that the uncertainty in $\Omega_{\eta+\eta'}$~\cite{omega}
affects the final
estimate only marginally since any change in this input implies a
change in the fit to the $\Delta I=1/2$ rule with the net effect of
recovering the original prediction. In fact, any change in
$\Omega_{\eta+\eta'}$ is
 anti-correlated to the value of \GG\ oobtained in the fit, the
variation of
which compensates in \ratio\ the original change of
$\Omega_{\eta+\eta'}$.
Changing the value of $m_c$ affects the value of \ratio\
 below the 10\% level. 

The value of \ratio\ depends---besides other given parameters
like meson masses and decay constants, and the matching scale
$\mu$ that we keep fixed---on eight input parameters:
M, \qq, \GG, $m_s$, $\Lambda_{\rm QCD}$, $m_t$, $V_{cb}$ and $\bar \rho$. 
The range for $\bar \rho$
is that obtained in the unitarity triangle without the bounds from the
experimental value of \eps. 
See Table I for a summary of the ranges used.
For central values of all input parameters, \ratio\ is equal to 
$2.5 \times 10^{-3}$ in both $\gamma_5$-schemes. This value
corresponds to the determination $\hat B_K=1.0$. By comparison with
the experimental value of \eps, it is then found that
$\Im \lambda_t = 1.3 \times  10^{-4}$ and $ 1.2 \times  10^{-4}$,
respectively, in the HV and NDR scheme.

\vbox{
\begin{table}
\begin{center}
\begin{footnotesize}
\begin{tabular}{c c}
{\rm parameter} & {\rm value} \\
\hline
$\Lambda^{(4)}_{\rm QCD}$ &  $340 \pm 40$ MeV\\
$ |V_{cb}|$ & $0.0405 \pm 0.0015$ \cite{stocchi}  \\
$\bar \rho$ & $0.242^{+0.070}_{-0.045}$ \cite{stocchi}\\
$\overline m_t (m_t)$ & $165 \pm 5$ GeV \\
$\Omega_{\eta+\eta'}$ & $0.25\pm 0.10$ \\
$\overline{m}_s$ (1 GeV) &  $150 \pm 25$ MeV \\
$\vev{\bar{q}q}$  & HV:  $( -210  \leftrightarrow -260 \:
\mbox{MeV} )^3$, NDR: $( -230 \leftrightarrow -260\:
\mbox{MeV} )^3$ \\
$ \langle \alpha_s GG/\pi \rangle $ &  HV: $(325 \leftrightarrow 335 \: 
\mbox{MeV} )^4 $, NDR: $(327 \leftrightarrow 340 \: 
\mbox{MeV} )^4 $ \\
$M$ & HV: $180 \leftrightarrow 220$ MeV, NDR:  $185 \leftrightarrow
210$ MeV\\
\end{tabular}
\end{footnotesize}
\end{center}
\caption{Table of the numerical values of the most
relevant input
parameters used in the present analysis. }
\label{inputs}
\end{table}
}

In order to estimate the error,
a set of values for \ratio\ is generated by varying the parameters
$M$, \GG, \qq\ in a flat scanning of their given ranges,  
while the others according to a normal distribution. 
The difference in the values
of \ratio\ found in the HV and NDR schemes is rather small in the
$\chi$QM approach and it can be considered as part of the overall
theoretical uncertainty by joining the values found in the two schemes.
The set of values of \ratio\ thus found  is amenable
of statistical analysis. For a given set, a distribution is
obtained by collecting the values of \ratio\ in bins of
a given range. This is shown in  Fig.~\ref{dists}.
The final distribution is
partially skewed, with more values closer to the lower end but
a longer tail toward larger values.

\vbox{\begin{figure}           
\begin{center}
\includegraphics[scale=0.5]{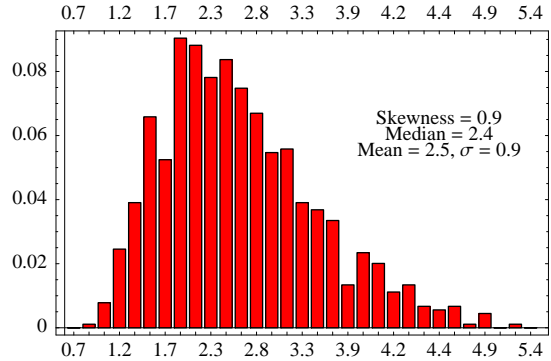}
\end{center}
\caption{Distribution of values of \ratio\ . Normalized bins versus
the central values ($\times 10^3$) of each bin.}
\label{dists}
\end{figure}}

However, because the skewness in the distribution is less than one
(and the standard deviation of the variance much smaller than the
variance itself),
the mean and the standard deviation are a good estimate of the central
value and the dispersion of values around it.
The statistical analysis therefore yields 
\be
\varepsilon '/\varepsilon = (2.5 \pm 0.9) \times 10^{-3}
\, , \label{stretto}
\ee
which is the best estimate quoted in the abstract.

The $1\sigma$ uncertainty of the estimate (\ref{stretto}) should be
considered together with the range given by the largest and the
smallest values of \ratio\ when all input parameters are varied
in a flat scanning within their errors.
In this case, it is found that 
\be
0.9 \times 10^{-3} < \varepsilon '/\varepsilon < 5.8 \times 10^{-3}
\, , \label{largo}
\ee
which is the more conservative result.

It is also possible to gain some insight on the dependence of
these results on the input parameters. For instance, in
Fig.~\ref{vars},
the generated values of \ratio\ are plotted against, respectively, \qq\ and
$m_s$ while randomly varying the other parameters. 
In particular, larger values of \qq\ correspond to larger
values of \ratio.

There is no strong dependence on $m_s$
or \GG\ because $m_s$ only enters in the NLO corrections and the 
gluon consensate does not contribute to the leading penguin operators. 
However, these two parameters  play a
role in the determination of \eps\ where the final values of $\hat
B_K$ depends on $m_s$ and \GG. Given (\ref{eps}), \ratio\  is
inversely proportional to the value of $\hat B_K$.

\vbox{\begin{figure}           
\begin{center}
\includegraphics[scale=0.5]{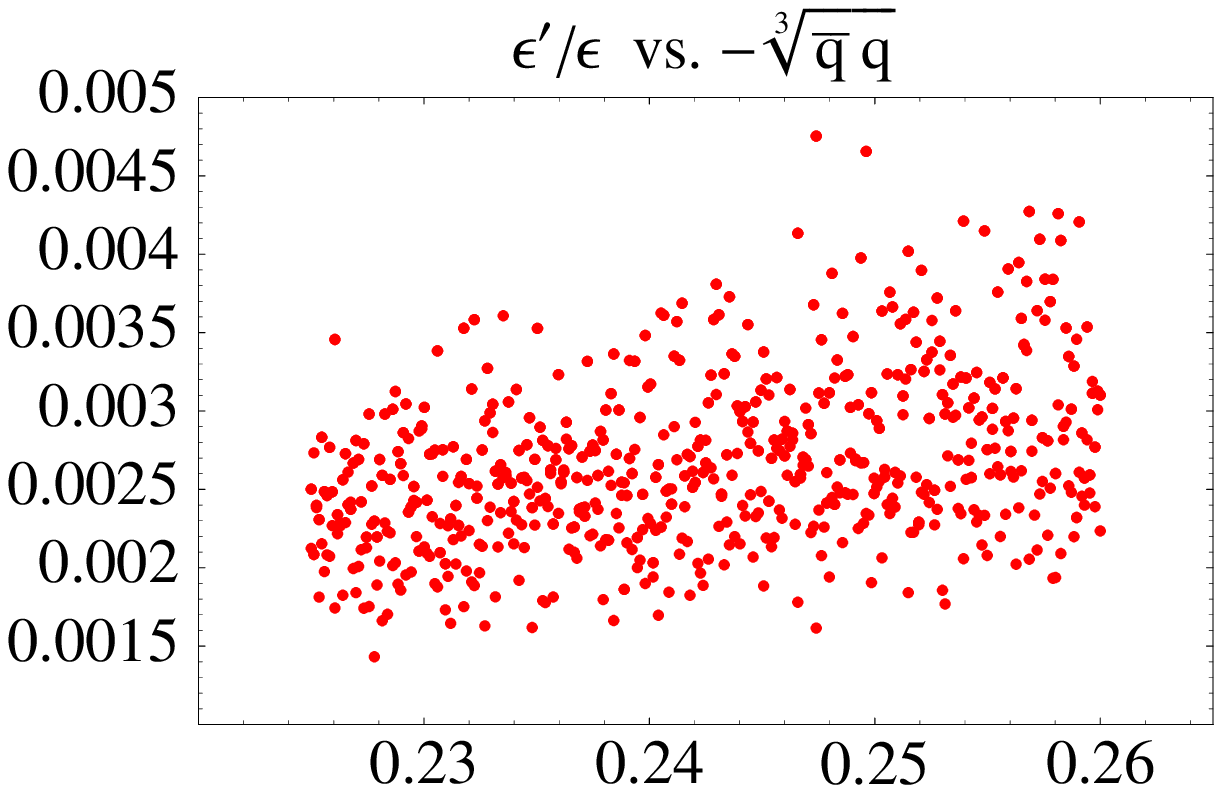}
\includegraphics[scale=0.5]{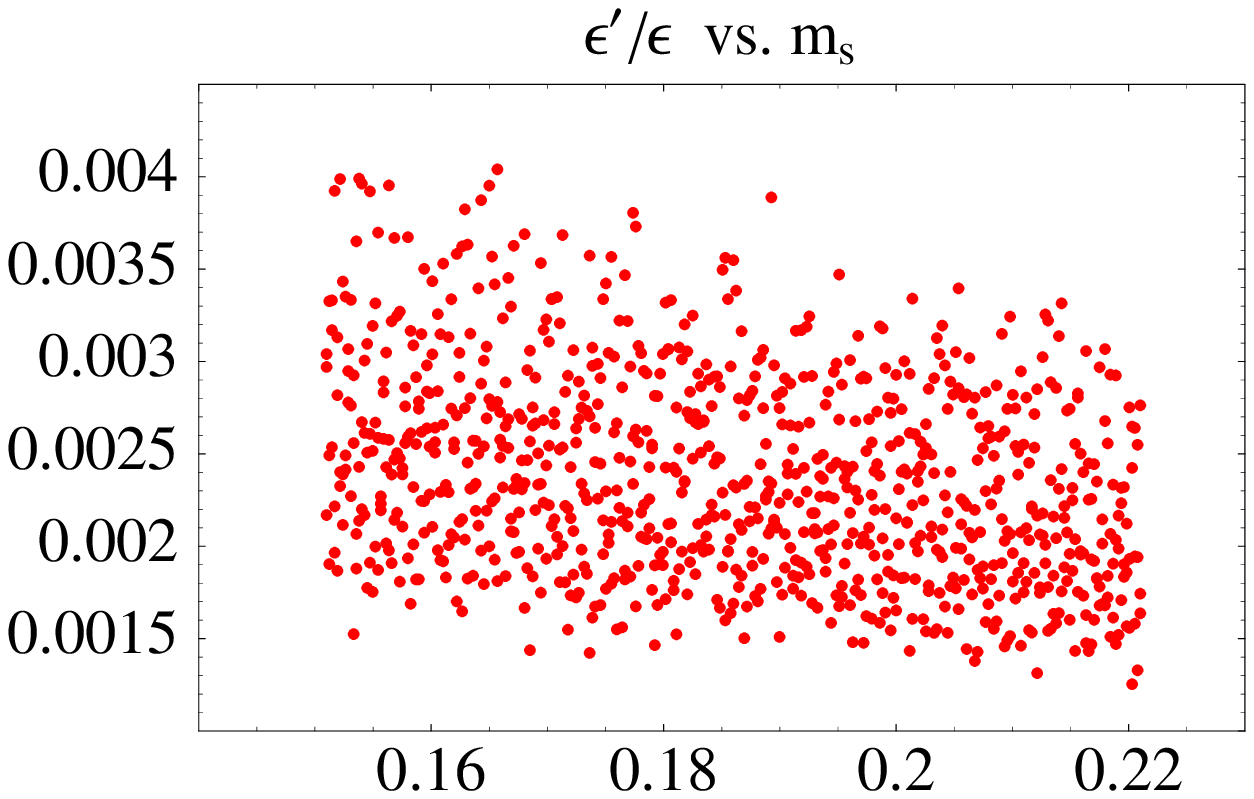}
\end{center}
\caption{Variations of \ratio\ as the quark condensate \qq\ and $m_s$
 are varied. Other inputs are randomly generated within their bounds.}
\label{vars}
\end{figure}}

 The results in
(\ref{stretto}) and (\ref{largo}) are depicted in 
Fig.~\ref{expvsQM} as yellow/light-gray
 and gray bands together with the most recent
experimental determinations and the average in (\ref{average}).
As it can be seen from this figure,
the result is in good agreement with the experimental data.

\vbox{\begin{figure}              
\begin{center}
\includegraphics[scale=0.6]{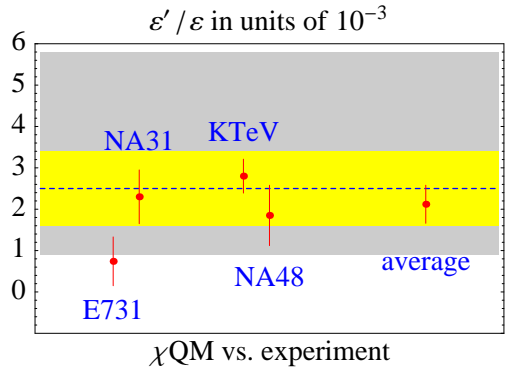}
\end{center}
\caption{Experiments vs.\ the present  estimate of \ratio.
 The dashed line is the
central value, the gray area spans the conservative range  (\ref{largo}) and
the yellow/light-gray  area spans
the 1$\sigma$ range in (\ref{stretto}).}
\label{expvsQM}
\end{figure}}

As discussed in more details somewhere else~\cite{nuovo}, the rather
large values of \ratio\ found by means of the $\chi$QM hadronic matrix
elements, and the corresponding better agreement with the experiments,
comes about because the fit of the $\Delta I=1/2$ rule---on which the
approach is based---enhances the contribution of the isospin $I=0$
channel which, in turns, 
drives \ratio\ toward larger values.
  
\vspace*{0.6cm}

I thank my collaborators S.\ Bertolini and
J.\ Eeg for discussions and F.\ Parodi and A.\ Stocchi
 for their help in the determination of $\bar \rho$.



\begin{references}

\bibitem{W}{\sc  L. Wolfenstein}, Phys. Rev. Lett. 13 (1964) 562.

\bibitem{KTeV} {\sc KTeV Collaboration (A. Alavi-Harati et~al.)}, 
Phys. Rev. Lett. 83
(1999) 22.

\bibitem{NA48} {\sc NA48 Collaboration (V. Fanti et al.)}, 
 Phys. Lett. B465 (1999) 335 and A. Ceccucci's talk, CERN February 29, 2000 
({\tt http://www.cern.ch/NA48/Welcome.html}).

\bibitem{NA31} {\sc NA31 Collaboration (G.~D.~Barr et~al.)}, 
Phys. Lett. B317 (1993) 233.

\bibitem{E731} {\sc E731 Collaboration (L.~K. Gibbons et~al.)},
Phys. Rev. D55 (1997) 6625.

\bibitem{WW}{\sc B. Winstein and L. Wolfenstein}, Rev. Mod. Phys. 
65 (1993) 1113.

\bibitem{burasrev} {\sc G. Buchalla and A. J. Buras and M. E.
 Lautenbacher}, Rev. Mod. Phys. 68 (1996) 1125.

\bibitem{review} {\sc S. Bertolini, J. Eeg and M. Fabbrichesi},
Rev. Mod. Phys. 72 (2000) 65.

\bibitem{Maiani} {\sc L. Maiani and G. Pancheri and N. Paver},
{\em The DAPHNE Physics Handbook} (Frascati, 1992).

\bibitem{NLO} {\sc A.~J.~Buras et al}, Nucl. Phys. B370 (1992) 69 and 
 Nucl. Phys. B400 (1993) 37;\\
{\sc A.~J.~Buras, M. Jamin and M.E. Lautenbacher},  Nucl. Phys. B400
(1993) 75 and  Nucl. Phys. B408 (1993) 209;\\
{\sc M. Ciuchini, E. Franco, G. Martinelli and L. Reina}, 
 Nucl. Phys. B415 (1994) 403.

\bibitem{IL} {\sc T. Inami and C. S. Lim}, Prog. Theor. Phys. 65
(1981) 297.

\bibitem{s2} {\sc A. J. Buras and M. Jamin and P. H. Weisz}, 
Nucl. Phys. B347 (1990) 491;\\
{\sc S. Herrlich and U. Nierste}, Phys. Rev. D52 (1995) 6505 and
Nucl. Phys. B476 (1996) 27.
  
\bibitem{dortmund} {\sc T. Hambye et al.}, {\tt hep-ph/9906434}.

\bibitem{dubna} {\sc A.~A. Bel'kov et al. },  {\tt hep-ph/9907335}.

\bibitem{taipei} {\sc H.-Y. Cheng}, {\tt hep-ph/9911202}.

\bibitem{munich99} {\sc S. Bosch et al.}, {\tt hep-ph/9904408}.

\bibitem{roma99}  {\sc M. Ciuchini \etal}, {\tt hep-th/9910236}.

\bibitem{trieste97} {\sc  S. Bertolini, J. Eeg and M. Fabbrichesi}, Nucl. Phys. B514 (1998)
63 and 93.

\bibitem{book} {\sc G.~C. Branco et al.}, CP Violation (Oxford
Science Publications, Oxford 1999). 

\bibitem{QM} {\sc K. Nishijima}, N. Cim. 11 (1959) 698;\\
{\sc F. Gursey},  N. Cim. 16 (1960) 23 and Ann. Phys. 12 (1961) 91;\\
{\sc J.A. Cronin}, Phys. Rev. 161 (1967 1483;\\
{\sc S. Weinberg}, Physica 96A (1979) 327;\\
{\sc A. Manohar and H. Georgi}, Nucl. Phys B234 (1984) 189;\\
{\sc A. Manohar and G. Moore},  Nucl. Phys B243 (1984) 55;\\
{\sc D. Espriu et al.},  Nucl. Phys B345 (1990) 22. 

\bibitem{pich}  {\sc A. Pich and E. de Rafael}, Nucl. Phys B358
(1991), 311.

\bibitem{nuovo} {\sc  S. Bertolini, J. Eeg and M. Fabbrichesi}, 
{\tt hep-ph/0002234}.

\bibitem{omega} For a recent discussion, see, e.g., {\sc S. Gardner
and G. Valencia},  Phys. Lett. B466 (1999) 355.

\bibitem{stocchi} {\sc F. Parodi, P. Roudeau and A. Stocchi}, 
N.\ Cim.\ A112 (1999) 833.
 
\end{references}
\end{document}